\documentclass[runningheads]{llncs}
\usepackage{algorithm,algpseudocode,cite,graphicx}
\usepackage[usenames,dvipsnames]{xcolor}

\newcommand{\MF}{\ensuremath{\mathrm{MF}}}
\newcommand{\MB}{\ensuremath{\mathrm{MB}}}
\newcommand{\LCP}{\ensuremath{\mathrm{LCP}}}
\newcommand{\LCS}{\ensuremath{\mathrm{LCS}}}
\newcommand{\search}{\ensuremath{\mathrm{search}}}
\newcommand{\FMT}{\ensuremath{\mathrm{FMT}}}
\newcommand{\FMrevT}{\ensuremath{\mathrm{FMT^{rev}}}}
\newcommand{\revP}{\ensuremath{P^\mathrm{rev}}}

\begin{document}

\title{How to Find Long Maximal Exact Matches\\and Ignore Short Ones}
\titlerunning{How to Find Long MEMs and Ignore Short Ones}

\author{Travis Gagie\orcidID{0000-0003-3689-327X}}
\authorrunning{T.\ Gagie}
\institute{Faculty of Computer Science\\Dalhousie University, Halifax, Canada\\
\email{travis.gagie@dal.ca}}

\maketitle

\begin{abstract}
Finding maximal exact matches (MEMs) between strings is an important task in bioinformatics, but it is becoming increasingly challenging as geneticists switch to pangenomic references.  Fortunately, we are usually interested only in the relatively few MEMs that are longer than we would expect by chance.  In this paper we show that under reasonable assumptions we can find all MEMs of length at least $L$ between a pattern of length $m$ and a text of length $n$ in $O (m)$ time plus extra $O (\log n)$ time only for each MEM of length at least nearly $L$ using a compact index for the text, suitable for pangenomics.

\keywords{Maximal exact matches \and pangenomics \and Burrows-Wheeler Transform \and grammar-based compression.}
\end{abstract}

\section{Introduction}
\label{sec:introduction}

Finding maximal exact matches (MEMs) has been an important task at least since Li's introduction of BWA-MEM~\cite{Li13}.  A MEM (in other contexts sometimes called a super-MEM or SMEM~\cite{Li12}) of a pattern $P [1..m]$ with respect to a text $T [1..n]$ is a non-empty substring $P [i..j]$ of $P$ such that
\begin{itemize}
\item $P [i..j]$ occurs in $T$,
\item $i = 1$ or $P [i - 1..j]$ does not occur in $T$,
\item $j = m$ or $P [i..j + 1]$ does not occur in $T$.
\end{itemize}
If we have a suffix tree for $T$ then we can find all the MEMs of $P$ with respect to $T$ in $O (m)$ time, but its $\Theta (n)$-word space bound is completely impractical in bioinformatics.  The textbook compact solution (see~\cite{Ohl13,Nav16,MBCT23}) uses a bidirectional FM-index to simulate a suffix tree, which is slightly slower --- typically using $\Theta (\log n)$ extra time per MEM, and with significantly worse constant coefficients overall --- but takes $\Theta (n)$ bits for DNA instead of $\Theta (n)$ words.  As geneticists have started aligning against pangenomic references consisting of hundreds or thousands of genomes, however, even $\Theta (n)$ bits is unacceptable.  Bioinformaticians have started designing indexes for MEM-finding that can work with such massive and highly repetitive datasets~\cite{GNP20,BGI20,Gao22,Nav23,KK23} but, although they have shown some practical promise~\cite{ROLGB22}, there is still definite room for improvement.

Fortunately, we are usually interested only in the relatively few MEMs that are longer than we would expect by chance.  For example, consider the randomly chosen string over $\{\mathtt{A}, \mathtt{C}, \mathtt{G}, \mathtt{T}\}$ shown at the top of Figure~\ref{fig:random}, with the highlighted substring copied below it and then edited by having each of its characters replaced with probability $1 / 4$ by another character chosen uniformly at random from $\{\mathtt{A}, \mathtt{C}, \mathtt{G}, \mathtt{T}\}$ (so a character could be replaced by a copy of itself).  The differences from the original substring are shown highlighted in the copy, with the lengths of the MEMs of the copy with respect to the whole string shown under the copy.  The occurrences of the MEMs in the whole string are shown at the bottom of the figure, with the two reasonably long MEMs --- of length 12 and 8 --- highlighted in red.  These are the two interesting MEMs, and the others are really more trouble than they are worth.  (The unhighlighted substrings with more than 6 characters are formed by consecutive or overlapping occurrences of MEMs with at most 6 characters.)  The whole string contains 225 of 256 possible distinct 4-tuples (88\%), 421 of 1024 possible distinct 5-tuples (41\%), and 512 of 4096 possible distinct 6-tuples (13\%), so even if the copied substring were completely scrambled we would still expect quite a lot of MEMs of these lengths --- and we should ignore them, since they are mostly just noise.

\begin{figure}[t!]
\begin{center}
{\tt \begin{tabular}{c}
TCTTAGCTGACGTTCGGGGCGGGTTAGGCCATCTTCTATAGATTTCTCAG \\
AGACATCCTAGCCGTGCTGAAGTTGTCACTCGCGGCCGTGTTTCCTAACG \\
CCACCTGATAGCGTGTTCCAAGCACTTGAGTGTCGGGCTGTAGGGGCTCA \\
CTCTGCGCAGGATCACGGCTGTTTGTACCTATATCGTTATCGTACTGAAT \\
\textcolor{red}{AAGTAGAATATCCAAACTTTCAGATTCCGGTTTGGCTGCCAAAACTAGGT} \\
GGGATGTGATGCGCGGCGAATTGTGATCTCGCATTGTATATTATCAATCT \\
CAGCTTAGCTTGACTTGCACAAAATGAACCCTACGGCGGTGGAGGATTAC \\
GACCGGAAGCGTCCTGCCTCGGAAAGCGTCCTCCTCAGAAGACGCGCGTG \\
AGGTCCGTCTTGTGGTCGCGACACAATACGCGACACGAACGACTGGTACC \\
GGATCAAGTTCTCGATAGGCTGAATTGGCTCTTGTATACATGATGATTGT \\
GGAATCTATACTGTGAACTTATAGGCAAATCCTATGCCACTACATTACGG \\[3ex]
AAGT\textcolor{red}{CTT}ATA\textcolor{red}{C}CCAAACTT\textcolor{red}{A}C\textcolor{red}{G}GATTCCGGTTTG\textcolor{red}{T}CTGCC\textcolor{red}{G}AAA\textcolor{red}{T}TAGGT \\
4\ 556\ 5\ 44\ 8\ \ 6\ \ 6\ \ 55(12)\ \ \ 6\ 5\ 455\ 4\ 4444455\ \ \ \ \mbox{} \\[3ex]
\textcolor{black}{T}\textcolor{black}{C}\textcolor{black}{T}\textcolor{black}{T}\textcolor{black}{A}\textcolor{lightgray}{G}\textcolor{lightgray}{C}\textcolor{lightgray}{T}\textcolor{lightgray}{G}\textcolor{lightgray}{A}\textcolor{lightgray}{C}\textcolor{lightgray}{G}\textcolor{lightgray}{T}\textcolor{lightgray}{T}\textcolor{lightgray}{C}\textcolor{lightgray}{G}\textcolor{lightgray}{G}\textcolor{lightgray}{G}\textcolor{lightgray}{G}\textcolor{lightgray}{C}\textcolor{lightgray}{G}\textcolor{lightgray}{G}\textcolor{lightgray}{G}\textcolor{black}{T}\textcolor{black}{T}\textcolor{black}{A}\textcolor{black}{G}\textcolor{black}{G}\textcolor{lightgray}{C}\textcolor{lightgray}{C}\textcolor{lightgray}{A}\textcolor{lightgray}{T}\textcolor{lightgray}{C}\textcolor{lightgray}{T}\textcolor{lightgray}{T}\textcolor{lightgray}{C}\textcolor{lightgray}{T}\textcolor{lightgray}{A}\textcolor{lightgray}{T}\textcolor{lightgray}{A}\textcolor{lightgray}{G}\textcolor{lightgray}{A}\textcolor{lightgray}{T}\textcolor{lightgray}{T}\textcolor{lightgray}{T}\textcolor{lightgray}{C}\textcolor{lightgray}{T}\textcolor{lightgray}{C}\textcolor{lightgray}{A}\textcolor{lightgray}{G} \\
\textcolor{lightgray}{A}\textcolor{lightgray}{G}\textcolor{lightgray}{A}\textcolor{lightgray}{C}\textcolor{lightgray}{A}\textcolor{lightgray}{T}\textcolor{lightgray}{C}\textcolor{lightgray}{C}\textcolor{lightgray}{T}\textcolor{lightgray}{A}\textcolor{black}{G}\textcolor{black}{C}\textcolor{black}{C}\textcolor{black}{G}\textcolor{lightgray}{T}\textcolor{lightgray}{G}\textcolor{lightgray}{C}\textcolor{lightgray}{T}\textcolor{lightgray}{G}\textcolor{black}{A}\textcolor{black}{A}\textcolor{black}{G}\textcolor{black}{T}\textcolor{black}{T}\textcolor{black}{G}\textcolor{black}{T}\textcolor{black}{C}\textcolor{lightgray}{A}\textcolor{lightgray}{C}\textcolor{lightgray}{T}\textcolor{lightgray}{C}\textcolor{lightgray}{G}\textcolor{lightgray}{C}\textcolor{lightgray}{G}\textcolor{black}{G}\textcolor{black}{C}\textcolor{black}{C}\textcolor{black}{G}\textcolor{lightgray}{T}\textcolor{lightgray}{G}\textcolor{lightgray}{T}\textcolor{lightgray}{T}\textcolor{lightgray}{T}\textcolor{lightgray}{C}\textcolor{lightgray}{C}\textcolor{lightgray}{T}\textcolor{lightgray}{A}\textcolor{lightgray}{A}\textcolor{lightgray}{C}\textcolor{lightgray}{G} \\
\textcolor{lightgray}{C}\textcolor{lightgray}{C}\textcolor{lightgray}{A}\textcolor{lightgray}{C}\textcolor{lightgray}{C}\textcolor{lightgray}{T}\textcolor{lightgray}{G}\textcolor{lightgray}{A}\textcolor{lightgray}{T}\textcolor{lightgray}{A}\textcolor{lightgray}{G}\textcolor{lightgray}{C}\textcolor{lightgray}{G}\textcolor{lightgray}{T}\textcolor{lightgray}{G}\textcolor{lightgray}{T}\textcolor{lightgray}{T}\textcolor{lightgray}{C}\textcolor{lightgray}{C}\textcolor{lightgray}{A}\textcolor{lightgray}{A}\textcolor{lightgray}{G}\textcolor{lightgray}{C}\textcolor{lightgray}{A}\textcolor{lightgray}{C}\textcolor{lightgray}{T}\textcolor{lightgray}{T}\textcolor{lightgray}{G}\textcolor{lightgray}{A}\textcolor{lightgray}{G}\textcolor{lightgray}{T}\textcolor{lightgray}{G}\textcolor{lightgray}{T}\textcolor{lightgray}{C}\textcolor{lightgray}{G}\textcolor{lightgray}{G}\textcolor{lightgray}{G}\textcolor{lightgray}{C}\textcolor{lightgray}{T}\textcolor{lightgray}{G}\textcolor{lightgray}{T}\textcolor{lightgray}{A}\textcolor{lightgray}{G}\textcolor{lightgray}{G}\textcolor{lightgray}{G}\textcolor{lightgray}{G}\textcolor{lightgray}{C}\textcolor{lightgray}{T}\textcolor{lightgray}{C}\textcolor{lightgray}{A} \\
\textcolor{lightgray}{C}\textcolor{black}{T}\textcolor{black}{C}\textcolor{black}{T}\textcolor{black}{G}\textcolor{black}{C}\textcolor{lightgray}{G}\textcolor{lightgray}{C}\textcolor{lightgray}{A}\textcolor{lightgray}{G}\textcolor{lightgray}{G}\textcolor{lightgray}{A}\textcolor{lightgray}{T}\textcolor{lightgray}{C}\textcolor{black}{A}\textcolor{black}{C}\textcolor{black}{G}\textcolor{black}{G}\textcolor{lightgray}{C}\textcolor{lightgray}{T}\textcolor{black}{G}\textcolor{black}{T}\textcolor{black}{T}\textcolor{black}{T}\textcolor{black}{G}\textcolor{black}{T}\textcolor{black}{A}\textcolor{black}{C}\textcolor{black}{C}\textcolor{lightgray}{T}\textcolor{lightgray}{A}\textcolor{lightgray}{T}\textcolor{lightgray}{A}\textcolor{lightgray}{T}\textcolor{lightgray}{C}\textcolor{lightgray}{G}\textcolor{lightgray}{T}\textcolor{lightgray}{T}\textcolor{lightgray}{A}\textcolor{lightgray}{T}\textcolor{lightgray}{C}\textcolor{lightgray}{G}\textcolor{lightgray}{T}\textcolor{lightgray}{A}\textcolor{lightgray}{C}\textcolor{lightgray}{T}\textcolor{lightgray}{G}\textcolor{lightgray}{A}\textcolor{lightgray}{A}\textcolor{lightgray}{T} \\
\textcolor{black}{A}\textcolor{black}{A}\textcolor{black}{G}\textcolor{black}{T}\textcolor{lightgray}{A}\textcolor{lightgray}{G}\textcolor{lightgray}{A}\textcolor{lightgray}{A}\textcolor{lightgray}{T}\textcolor{lightgray}{A}\textcolor{lightgray}{T}\textcolor{red}{CCAAACTT}\textcolor{lightgray}{T}\textcolor{lightgray}{C}\textcolor{lightgray}{A}\textcolor{red}{GATTCCGGTTTG}\textcolor{lightgray}{G}\textcolor{black}{C}\textcolor{black}{T}\textcolor{black}{G}\textcolor{black}{C}\textcolor{black}{C}\textcolor{lightgray}{A}\textcolor{lightgray}{A}\textcolor{lightgray}{A}\textcolor{lightgray}{A}\textcolor{lightgray}{C}\textcolor{black}{T}\textcolor{black}{A}\textcolor{black}{G}\textcolor{black}{G}\textcolor{black}{T} \\
\textcolor{lightgray}{G}\textcolor{lightgray}{G}\textcolor{lightgray}{G}\textcolor{lightgray}{A}\textcolor{lightgray}{T}\textcolor{lightgray}{G}\textcolor{lightgray}{T}\textcolor{lightgray}{G}\textcolor{lightgray}{A}\textcolor{lightgray}{T}\textcolor{lightgray}{G}\textcolor{lightgray}{C}\textcolor{lightgray}{G}\textcolor{lightgray}{C}\textcolor{lightgray}{G}\textcolor{lightgray}{G}\textcolor{black}{C}\textcolor{black}{G}\textcolor{black}{A}\textcolor{black}{A}\textcolor{black}{T}\textcolor{black}{T}\textcolor{lightgray}{G}\textcolor{lightgray}{T}\textcolor{lightgray}{G}\textcolor{lightgray}{A}\textcolor{lightgray}{T}\textcolor{lightgray}{C}\textcolor{lightgray}{T}\textcolor{lightgray}{C}\textcolor{lightgray}{G}\textcolor{lightgray}{C}\textcolor{lightgray}{A}\textcolor{lightgray}{T}\textcolor{lightgray}{T}\textcolor{lightgray}{G}\textcolor{lightgray}{T}\textcolor{lightgray}{A}\textcolor{lightgray}{T}\textcolor{black}{A}\textcolor{black}{T}\textcolor{black}{T}\textcolor{black}{A}\textcolor{lightgray}{T}\textcolor{lightgray}{C}\textcolor{lightgray}{A}\textcolor{lightgray}{A}\textcolor{lightgray}{T}\textcolor{lightgray}{C}\textcolor{lightgray}{T} \\
\textcolor{lightgray}{C}\textcolor{lightgray}{A}\textcolor{lightgray}{G}\textcolor{black}{C}\textcolor{black}{T}\textcolor{black}{T}\textcolor{black}{A}\textcolor{lightgray}{G}\textcolor{lightgray}{C}\textcolor{lightgray}{T}\textcolor{lightgray}{T}\textcolor{lightgray}{G}\textcolor{lightgray}{A}\textcolor{lightgray}{C}\textcolor{lightgray}{T}\textcolor{lightgray}{T}\textcolor{lightgray}{G}\textcolor{lightgray}{C}\textcolor{lightgray}{A}\textcolor{lightgray}{C}\textcolor{lightgray}{A}\textcolor{black}{A}\textcolor{black}{A}\textcolor{black}{A}\textcolor{black}{T}\textcolor{lightgray}{G}\textcolor{lightgray}{A}\textcolor{black}{A}\textcolor{black}{C}\textcolor{black}{C}\textcolor{black}{C}\textcolor{lightgray}{T}\textcolor{black}{A}\textcolor{black}{C}\textcolor{black}{G}\textcolor{black}{G}\textcolor{lightgray}{C}\textcolor{lightgray}{G}\textcolor{lightgray}{G}\textcolor{lightgray}{T}\textcolor{lightgray}{G}\textcolor{lightgray}{G}\textcolor{lightgray}{A}\textcolor{black}{G}\textcolor{black}{G}\textcolor{black}{A}\textcolor{black}{T}\textcolor{black}{T}\textcolor{black}{A}\textcolor{lightgray}{C} \\
\textcolor{lightgray}{G}\textcolor{lightgray}{A}\textcolor{lightgray}{C}\textcolor{lightgray}{C}\textcolor{lightgray}{G}\textcolor{lightgray}{G}\textcolor{lightgray}{A}\textcolor{lightgray}{A}\textcolor{lightgray}{G}\textcolor{lightgray}{C}\textcolor{lightgray}{G}\textcolor{lightgray}{T}\textcolor{lightgray}{C}\textcolor{black}{C}\textcolor{black}{T}\textcolor{black}{G}\textcolor{black}{C}\textcolor{black}{C}\textcolor{lightgray}{T}\textcolor{lightgray}{C}\textcolor{lightgray}{G}\textcolor{black}{G}\textcolor{black}{A}\textcolor{black}{A}\textcolor{black}{A}\textcolor{lightgray}{G}\textcolor{lightgray}{C}\textcolor{lightgray}{G}\textcolor{lightgray}{T}\textcolor{lightgray}{C}\textcolor{lightgray}{C}\textcolor{lightgray}{T}\textcolor{lightgray}{C}\textcolor{lightgray}{C}\textcolor{lightgray}{T}\textcolor{lightgray}{C}\textcolor{lightgray}{A}\textcolor{lightgray}{G}\textcolor{lightgray}{A}\textcolor{lightgray}{A}\textcolor{lightgray}{G}\textcolor{lightgray}{A}\textcolor{lightgray}{C}\textcolor{lightgray}{G}\textcolor{lightgray}{C}\textcolor{lightgray}{G}\textcolor{lightgray}{C}\textcolor{lightgray}{G}\textcolor{lightgray}{T}\textcolor{lightgray}{G} \\
\textcolor{lightgray}{A}\textcolor{lightgray}{G}\textcolor{lightgray}{G}\textcolor{lightgray}{T}\textcolor{lightgray}{C}\textcolor{lightgray}{C}\textcolor{black}{G}\textcolor{black}{T}\textcolor{black}{C}\textcolor{black}{T}\textcolor{black}{T}\textcolor{lightgray}{G}\textcolor{lightgray}{T}\textcolor{lightgray}{G}\textcolor{lightgray}{G}\textcolor{lightgray}{T}\textcolor{lightgray}{C}\textcolor{lightgray}{G}\textcolor{lightgray}{C}\textcolor{lightgray}{G}\textcolor{lightgray}{A}\textcolor{lightgray}{C}\textcolor{lightgray}{A}\textcolor{lightgray}{C}\textcolor{lightgray}{A}\textcolor{lightgray}{A}\textcolor{lightgray}{T}\textcolor{lightgray}{A}\textcolor{lightgray}{C}\textcolor{lightgray}{G}\textcolor{lightgray}{C}\textcolor{lightgray}{G}\textcolor{lightgray}{A}\textcolor{lightgray}{C}\textcolor{lightgray}{A}\textcolor{black}{C}\textcolor{black}{G}\textcolor{black}{A}\textcolor{black}{A}\textcolor{lightgray}{C}\textcolor{lightgray}{G}\textcolor{lightgray}{A}\textcolor{lightgray}{C}\textcolor{lightgray}{T}\textcolor{lightgray}{G}\textcolor{lightgray}{G}\textcolor{black}{T}\textcolor{black}{A}\textcolor{black}{C}\textcolor{black}{C} \\
\textcolor{black}{G}\textcolor{black}{G}\textcolor{black}{A}\textcolor{black}{T}\textcolor{lightgray}{C}\textcolor{black}{A}\textcolor{black}{A}\textcolor{black}{G}\textcolor{black}{T}\textcolor{lightgray}{T}\textcolor{lightgray}{C}\textcolor{lightgray}{T}\textcolor{lightgray}{C}\textcolor{lightgray}{G}\textcolor{lightgray}{A}\textcolor{lightgray}{T}\textcolor{lightgray}{A}\textcolor{lightgray}{G}\textcolor{lightgray}{G}\textcolor{lightgray}{C}\textcolor{lightgray}{T}\textcolor{lightgray}{G}\textcolor{black}{A}\textcolor{black}{A}\textcolor{black}{T}\textcolor{black}{T}\textcolor{lightgray}{G}\textcolor{lightgray}{G}\textcolor{lightgray}{C}\textcolor{lightgray}{T}\textcolor{lightgray}{C}\textcolor{lightgray}{T}\textcolor{lightgray}{T}\textcolor{lightgray}{G}\textcolor{black}{T}\textcolor{black}{A}\textcolor{black}{T}\textcolor{black}{A}\textcolor{black}{C}\textcolor{lightgray}{A}\textcolor{lightgray}{T}\textcolor{lightgray}{G}\textcolor{lightgray}{A}\textcolor{lightgray}{T}\textcolor{lightgray}{G}\textcolor{lightgray}{A}\textcolor{lightgray}{T}\textcolor{lightgray}{T}\textcolor{lightgray}{G}\textcolor{lightgray}{T} \\
\textcolor{lightgray}{G}\textcolor{lightgray}{G}\textcolor{lightgray}{A}\textcolor{lightgray}{A}\textcolor{lightgray}{T}\textcolor{lightgray}{C}\textcolor{black}{T}\textcolor{black}{A}\textcolor{black}{T}\textcolor{black}{A}\textcolor{black}{C}\textcolor{lightgray}{T}\textcolor{lightgray}{G}\textcolor{lightgray}{T}\textcolor{lightgray}{G}\textcolor{black}{A}\textcolor{black}{A}\textcolor{black}{C}\textcolor{black}{T}\textcolor{black}{T}\textcolor{black}{A}\textcolor{black}{T}\textcolor{black}{A}\textcolor{lightgray}{G}\textcolor{lightgray}{G}\textcolor{lightgray}{C}\textcolor{black}{A}\textcolor{black}{A}\textcolor{black}{A}\textcolor{black}{T}\textcolor{lightgray}{C}\textcolor{lightgray}{C}\textcolor{lightgray}{T}\textcolor{lightgray}{A}\textcolor{lightgray}{T}\textcolor{lightgray}{G}\textcolor{lightgray}{C}\textcolor{lightgray}{C}\textcolor{lightgray}{A}\textcolor{lightgray}{C}\textcolor{lightgray}{T}\textcolor{lightgray}{A}\textcolor{lightgray}{C}\textcolor{black}{A}\textcolor{black}{T}\textcolor{black}{T}\textcolor{black}{A}\textcolor{black}{C}\textcolor{black}{G}\textcolor{black}{G}
\end{tabular}}
\caption{A randomly chosen string {\bf (top)} over $\{\mathtt{A}, \mathtt{C}, \mathtt{G}, \mathtt{T}\}$ with the highlighted substring copied {\bf (center)} and then edited.  The differences from the original substring are shown highlighted in red in the copy, with the lengths of the MEMs of the copy with respect to the whole string shown under the copy; 12 is shown as (12) to distinguish it from 1 followed by 2.  The occurrences of the MEMs in the whole string {\bf (bottom)} are shown in black when they have lengths 4, 5 or 6, and in red when they have lengths 8 or 12.  Substrings longer than 6 characters shown in black are formed by consecutive or overlapping occurrences of MEMs of length at most 6.}
\label{fig:random}
\end{center}
\end{figure}

In this paper we show how to find long, interesting MEMs without wasting time finding all the short, distracting ones.  We show that under reasonable assumptions we can find all the MEMs of length at least $L$ in time $O (m)$ time plus extra $O (\log n)$ time only for each MEM of length at least nearly $L$, using a compact index suitable for pangenomics.  Specifically, suppose the size of the alphabet is polylogarithmic in $n$, $\epsilon$ is a constant strictly between 0 and 1, $L \in \Omega (\log n)$ and we are given a straight-line program with $g$ rules for $T$.  Then there is an $O (r + \bar{r} + g)$-space index for $T$, where $r$ and $\bar{r}$ are the numbers of runs in the Burrows-Wheeler Transforms of $T$ and of the reverse of $T$, with which when given $P$ we can find all the MEMs of $P$ with respect to $T$ with length at least $L$ correctly with high probability and in $O (m + \mu_{(1 - \epsilon) L} \log n)$ time, where $\mu_x$ is the number of MEMs of length at least $x$.

The closest previous work to this paper is Li's~\cite{Li12} forward-backward algorithm for finding all MEMs, and a recent paper by Goga et al.'s~\cite{GDBFGN24} about how lazy evaluation of longest common prefix (LCP) queries can speed up finding long MEMs in practice.  We review those results and some related background in Section~\ref{sec:previous} and then combine them in Section~\ref{sec:result} to obtain our result.  In the appendix we present some preliminary experimental results.

\section{Previous Work}
\label{sec:previous}

As far as we know, the asymptotically fastest way to compute the MEMs of a pattern $P [1..m]$ with respect to an indexed text $T [1..n]$ using one of the indexes designed for massive and highly repetitive datasets, is to first compute the forward-match and backward-match pointers of $P$ with respect to $T$.  Figure~\ref{fig:MFMB} shows a small example of match pointers.

\begin{definition}
Let $\MF [1..m]$ and $\MB [1..m]$ be arrays of positions in $T$ such that $T [\MF [i]..n]$ has the longest common prefix with $P [i..m]$ of any suffix of $T$ and $T [1..\MB [i]]$ has the longest common suffix with $P [1..i]$ of any prefix of $T$, for $1 \leq i \leq m$.  We call $\MF [1..m]$ and $\MB [1..m]$ the {\em forward-match} and {\em back-ward match pointers} of $P$ with respect to $T$.
\end{definition}

\begin{figure}[t]
\begin{center}
\begin{tabular}{rrrrrrrrrrrrr}
        &     1 &     2 &     3 &     4 &     \textcolor{red}{5} &     6 &     7 &     8 &     9 &    10 &    11 &    12 \\
\hline
$T =$   & \tt G & \tt A & \tt T & \tt T & \tt \textcolor{red}{A} & \tt \textcolor{red}{G} & \tt \textcolor{red}{A} & \tt \textcolor{red}{T} & \tt A & \tt \textcolor{ProcessBlue}{C} & \tt \textcolor{ProcessBlue}{A} & \tt \textcolor{ProcessBlue}{T} \\
$P =$   & \tt T & \tt A & \tt \textcolor{ProcessBlue}{C} & \tt \textcolor{ProcessBlue}{A} & \tt \textcolor{ProcessBlue}{T} & \tt \textcolor{red}{A} & \tt \textcolor{red}{G} & \tt \textcolor{red}{A} & \tt \textcolor{red}{T} & \tt T & \tt A & \tt G \\
\hline
$\MF =$ &     8 &     9 &    10 &     7 &     4 &     \textcolor{red}{5} &     1 &     2 &     3 &     4 &     5 &     1 \\
$\MB =$ &     3 &     5 &    10 &    11 &    \textcolor{ProcessBlue}{12} &     9 &     6 &     7 &     8 &     4 &     5 &     6
\end{tabular}
\caption{The forward-match and backward-match pointers $\MF [1..m]$ and $\MB [1..m]$ of $P = \mathtt{TACATAGATTAG}$ with respect to $T = \mathtt{GATTAGATACAT}$.  Since $T [5..12]$ has the longest common prefix {\tt AGAT} with $P [6..12]$, $\MF [6] = 5$ {\bf (red)}; since $T [1..12]$ has the longest common suffix {\tt CAT} with $P [1..5]$, $\MB [5] = 12$ {\bf (blue)}.}
\label{fig:MFMB}
\end{center}
\end{figure}

Bannai, Gagie and I~\cite{BGI20} showed how to compute $\MF$ in $O (m (\log \log n + \log \sigma))$ time using an $O (r)$-space index for $T$, where $\sigma$ is the size of the alphabet and $r$ is the number of runs in the Burrows-Wheeler Transform (BWT) of $T$.  Applying a speedup by Nishimoto and Tabei~\cite{NT21}, their time bound becomes $O (m \log \sigma)$, or $O (m)$ when $\sigma$ is polylogarithmic in $n$ (using multiary wavelet trees~\cite{FMMN07}).  If we apply the same ideas to the reverses of $P$ and $T$, we can compute $\MB$ in the same time with an $O (\bar{r})$-space index, where $\bar{r}$ is the number of runs in the BWT of the reverse of $T$.

\begin{theorem}
\label{thm:pointers}
There is an $O (r + \bar{r})$-space index for $T$, where $r$ and $\bar{r}$ are the number of runs in the BWT of $T$ and the reverse of $T$, with which when given $P$ we can compute $\MF$ and $\MB$ in $O (m \log \sigma)$ time, or $O (m)$ time when $\sigma$ is polylogarithmic in $n$.
\end{theorem}

Suppose we have $\MF$ and $\MB$ and we can compute in $O (t (n))$ time both the length $\LCP \left( \rule{0ex}{2ex} P [i..m], T [\MF [i]..n] \right)$ of the longest common prefix of $P [i..m]$ and $T [\MF [i]..n]$ and the length $\LCS \left( \rule{0ex}{2ex} P [1..i], T [1..\MB [i]] \right)$ of the longest common suffix of $P [1..i]$ and $T [1..\MB [i]]$.  Then we can use a version of Li's~\cite{Li12} forward-backward algorithm to find all MEMs.  To see why, suppose we know that the $k$th MEM from the left starts at $P [i_k]$; then it ends at $P [j_k]$, where
\[j_k = i_k + \LCP \left( \rule{0ex}{2ex} P [i_k..m], T [\MF [i_k]..n] \right) - 1\,,\]
which we can find in $O (t (n))$ time.  Since MEMs cannot nest, the next character $P [j_k + 1]$ is in the $(k + 1)$st MEM from the left.  (For simplicity and without loss of generality, we assume all the characters in $P$ occur in $T$; otherwise, since MEMs cannot cross characters that do not occur in $T$, we split $P$ into maximal subpatterns consisting only of characters that do.)  That MEM starts at $P [i_{k + 1}]$, where
\[i_{k + 1} = (j_k + 1) - \LCS \left( \rule{0ex}{2ex} P [1..j_k + 1], T [1..\MB [j_k + 1]] \right) + 1\,,\]
which we can also find in $O (t (n))$ time.  If there are $\mu$ MEMs then, since the first from the left starts at $P [1]$, we can find them all in $O (\mu t (n))$ time.

Suppose we are given a straight-line program with $g$ rules for $T$.  By balancing it~\cite{GJL21} and augmenting its symbols with the Karp-Rabin hashes of their expansions, we can build an $O (g)$-space data structure with which, given $i$ and $j$ and constant-time access to the Karp-Rabin hashes of the substrings of $P$ --- which we can support after $O (m)$-time preprocessing of $P$ --- we can compute $\LCP (P [i..m], T [j..n])$ and $\LCS (P [1..i], T [1..j])$ correctly with high probability and in $O (\log n)$ time; see~\cite[Appendix A]{Dep24} for more details of the implementation.

Verbin and Yu~\cite{VY13} showed that any data structure using space $S$ polynomial in $g$ needs $\Omega \left( \frac{\log^{1 - \delta} n}{\log S} \right)$ time for random access to $T$ in the worst case, for any positive constant $\delta$, and Kempa and Kociumaka~\cite{KK22} showed that $r, \bar{r} \in O (g \log^2 n)$.  Since $g \in \Omega (\log n)$ and we can use our $\LCP$ or $\LCS$ queries to support random access to $T$ in $O (\sigma t (n))$ time, it follows that we cannot have $t (n)$ significantly sublogarithmic while still using space polynomial in $r + \bar{r} + g$ in the worst case.

Using our straight-line program to get $t (n) \in O (\log n)$ gives us the following result, which we believe to be the current state of the art.

\begin{theorem}
\label{thm:all}
There is an $O (r + \bar{r} + g)$ index for $T$ with which, given $P$, we can find all the $\mu$ MEMs of $P$ with respect to $T$ correctly with high probability and in $O (m \log \sigma + \mu \log n)$ time, or $O (m + \mu \log n)$ time when $\sigma$ is polylogarithmic in $n$.
\end{theorem}

Goga et al.~\cite{GDBFGN24} recently noted that if a MEM starts at $P [i]$ and $j$ is the next value at least $i + L - 1$ such that $j = m$ or $\MB [j + 1] \neq \MB [j] + 1$, then any MEM of length at least $L$ starting in $P [i..j]$ includes $P [j]$.  To see why, consider that a MEM of length at least $L$ starting after $P [i]$ must end at or after $P [i + L - 1]$, and that a MEM cannot end at $P [j]$ if $\MB [j + 1] = \MB [j] + 1$.  This means that if we are searching only for MEMs of length at least a given $L$ and we have performed an $\LCS$ query and found that a MEM starts at $P [i]$, then we can wait to perform another $\LCS$ query until we reach $P [j]$.

To see how much faster Goga et al.'s approach can be than finding all MEMs with Theorem~\ref{thm:all}, suppose $L \in \omega(\log n)$, the longest common substring of $P$ and $T$ has length $O (\log n)$, and there are $\Theta (m)$ MEMs.  Then when we evaluate $\LCS$ queries lazily we use only $O \left( m + \frac{m \log n}{L} \right) = O (m)$ time --- dominated by the time to compute $\MB$ --- but with Theorem~\ref{thm:all} we use $\Theta (m \log n)$ time.  (For consistency with the literature about the BWT and matching statistics, Goga et al.\ work right to left and so presented their approach as lazy $\LCP$ evaluation rather than lazy $\LCS$ evaluation.)  

On the other hand, if every proper prefix $P [1..i]$ of $P$ occurs in $T$, followed sometimes by $P [i + 1]$ and sometimes by some other character, then we could have $\MB [j + 1] \neq \MB [j] + 1$ for every $j < m$, even though every proper substring of $P$ can be extended to a longer match in $T$ and thus $P$ itself is the only MEM.  For example, if $P = \mathtt{GATTACAT}$ and \newline
\smallskip
\resizebox{\textwidth}{!}
{\small \tt \begin{tabular}{rrrrrrrrrrrrrrrrrrrrrrrrrrrrrrrrrrrrrrrrrrrr}
      & \tiny 1 & \tiny 2 & \tiny 3 & \tiny 4 & \tiny 5 & \tiny 6 & \tiny 7 & \tiny 8 & \tiny 9 & \tiny 10 & \tiny 11 & \tiny 12 & \tiny 13 & \tiny 14 & \tiny 15 & \tiny 16 & \tiny 17 & \tiny 18 & \tiny 19 & \tiny 20 & \tiny 21 & \tiny 22 & \tiny 23 & \tiny 24 & \tiny 25 & \tiny 26 & \tiny 27 & \tiny 28 & \tiny 29 & \tiny 30 & \tiny 31 & \tiny 32 & \tiny 33 & \tiny 34 & \tiny 35 & \tiny 36 & \tiny 37 & \tiny 38 & \tiny 39 & \tiny 40 & \tiny 41 & \tiny 42 & \tiny 43 \\
\hline
$T =$ & G & C & G & A & A & G & A & T & A & G & A & T & T & C & G & A & T & T & A & G & G & A & T & T & A & C & C & G & A & T & T & A & C & A & A & G & A & T & T & A & C & A & T \\
\end{tabular}}
\smallskip
then we can have $\MB [1..8] = [1, 4, 8, 13, 19, 26, 34, 43]$.  On an example like this, Goga et al.'s approach can use $\Omega (m \log n)$ time, while Theorem~\ref{thm:all} says we can use $O (m \log \sigma + \log n)$ time.

The weakness of Theorem~\ref{thm:all} is that it spends logarithmic time on each MEM, and the weakness of Goga et al.'s approach is that when it finds a very long MEM it can spend logarithmic time on each prefix longer than $L$ of that MEM.  When there are both many short MEMs and a few very long MEMs, neither approach may work well.

\section{Result}
\label{sec:result}

Suppose again that we have $\MF$ and $\MB$ and we can compute in $O (t (n))$ time both $\LCP \left( \rule{0ex}{2ex} P [i..m], T [\MF [i]..n] \right)$ and $\LCS \left( \rule{0ex}{2ex} P [1..i], T [1..\MB [i]] \right)$ for any $i$, and we are interested only in MEMs of length at least a given threshold $L$.  We now show how to modify the forward-backward algorithm to find only those MEMs, using something like Goga et al.'s approach.

Assume we have already found all MEMs of length at least $L$ that start in $P [1..i_k - 1]$ and that $P [i_k]$ is the start of a MEM, for some $i_k \leq m - L + 1$.  Notice that any MEMs of length at least $L$ that start in $P [i_k..i_k + L - 1]$ include $P [i_k + L - 1]$.  We set
\[b = \LCS \left( \rule{0ex}{2ex} P [1..i_k + L - 1], T [1..\MB [i_k + L - 1]] \right)\]
and consider two cases:
\begin{enumerate}

\item If $b \geq L$ then we set
\[f = \LCP \left( \rule{0ex}{2ex} P [i_k..m], T [\MF [i_k]..n] \right) \geq L\,,\]
so $P [i_k..i_k + f - 1]$ is the next MEM of length at least $L$.  We report $P [i_k..i_k + f - 1]$ and --- unless $i_k + f - 1 = m$ and we stop --- set $i_{k + 1}$ to the starting position
\[i_k + f - \LCS \left( \rule{0ex}{2ex} P [1..i_k + f], T [1..\MB [i_k + f] \right) + 1\]
of the next MEM from the left after $P [i_k..i_k + f - 1]$.

\item If $b < L$ then there is no MEM of length $L$ starting in $P [i_k..i_k + L - b - 1]$, so we set $i_{k + 1} = i_k + L - b$ which is the starting position of a MEM by our choice of $b$.

\end{enumerate}
After this, we have either reported all MEMs of length at least $L$ and stopped, or we have reported all MEMs of length at least $L$ that start in $P [1..i_{k + 1} - 1]$ with $i_{k + 1} > i_k$ and $P [i_{k + 1}]$ is the starting position of a MEM.  Some readers may wonder whether we can ever have $b > L$; notice that $P [i_k - 1]$ could be the start of one MEM of length much more than $L$ and $P [i_k]$ could be the start of another, in which case
\[\LCS \left( \rule{0ex}{2ex} P [1..i_k + L - 1], T [1..\MB [i_k + L - 1]] \right) \geq L + 1\,.\]

Algorithm~\ref{alg:BF} shows our pseudocode, starting with $i$ set to 1 and increasing it until it exceeds $m - L + 1$.  Figure~\ref{fig:trace} shows a trace of how Algorithm~\ref{alg:BF} processes our example of $P = \mathtt{TACATAGATTAG}$ and $T = \mathtt{GATTAGATACAT}$ from Figure~\ref{fig:MFMB}, with $L = 4$.  The reader may wonder why we do not follow Goga et al.\ more closely and set
\[b = \LCS \left( \rule{0ex}{2ex} P [1..j], T [1..\MB [j]] \right)\,,\]
where $j$ is the next value at least $i + L - 1$ such that $j = m$ or $\MB [j + 1] \neq \MB [j] + 1$, and adjust the rest of the algorithm accordingly.  This could indeed be faster in some cases but we do not see that our worst-case bounds (and it complicates our pseudocode and trace).

\begin{algorithm}[t]
\caption{Pseudocode for our version of Li's forward-backward algorithm, modified to find only MEMs of length at least $L$.}
\label{alg:BF}
\begin{algorithmic}[1]
\State{$i \gets 1$}
\While{$i \leq m - L + 1$}
  \State{$b \gets \LCS \left( \rule{0ex}{2ex} P [1..i + L - 1], T [1..\MB [i + L - 1]] \right)$}
  \If{$b \geq L$}
    \State{$f \gets \LCP \left( \rule{0ex}{2ex} P [i..m], T [\MF [i]..n] \right)$}
    \State{{\bf report} $P [i..i + f - 1]$}
    \If{$i + f - 1 = m$}
      \State{\bf break}
    \EndIf
    \State{$i \gets i + f - \LCS \left( \rule{0ex}{2ex} P [1..i + f], T [1..\MB [i + f]] \right) + 1$}
  \Else
    \State{$i \gets i + L - b$}
  \EndIf
\EndWhile
\end{algorithmic}
\end{algorithm}

\begin{figure}[p]
\begin{center}
\begin{tabular}{c@{\hspace{0.5ex}}r@{\hspace{1ex}}l}
line &  1: & $i \leftarrow 1$ \\[.5ex]
line &  2: & $i \leq 9$ \\[.5ex]
line &  3: & $\MB [4] = 11$ so $b \leftarrow \LCS (P [1..4], T [1..11]) = 4$ \\[.5ex]
line &  4: & $b \geq 4$ \\[.5ex]
line &  5: & $\MF [1] = 8$ so $f \leftarrow \LCP (P [1..12], T [8..12]) = 5$ \\[.5ex]
line &  6: & we report $P [1..5]$ \\[.5ex]
line &  7: & $i + f - 1 \neq 12$ \\[.5ex]
line & 10: & $\MB [6] = 9$ so $i \leftarrow 6 - \LCS (P [1..6], T [1..9]) + 1 = 4$ \\[2ex]
line &  2: & $i \leq 9$ \\[.5ex]
line &  3: & $\MB [7] = 6$ so $b \leftarrow \LCS (P [1..7], T [1..6]) = 3$ \\[.5ex]
line &  4: & $b < 4$ \\[.5ex]
line & 12: & $i \leftarrow 5$ \\[2ex]
line &  2: & $i \leq 9$ \\[.5ex]
line &  3: & $\MB [8] = 7$ so $b \leftarrow \LCS (P [1..8], T [1..7]) = 4$ \\[.5ex]
line &  4: & $b \geq 4$ \\[.5ex]
line &  5: & $\MF [5] = 4$ so $f \leftarrow \LCP (P [5..12], T [4..12]) = 5$ \\[.5ex]
line &  6: & we report $P [5..9]$ \\[.5ex]
line &  7: & $i + f - 1 \neq 12$ \\[.5ex]
line & 10: & $\MB [10] = 4$ so $i \leftarrow 10 - \LCS (P [1..10], T [1..4]) + 1 = 7$ \\[2ex]
line &  2: & $i \leq 9$ \\[.5ex]
line &  3: & $\MB [10] = 4$ so $b \leftarrow \LCS (P [1..10], T [1..4]) = 4$ \\[.5ex]
line &  4: & $b \geq 4$ \\[.5ex]
line &  5: & $\MF [7] = 1$ so $f \leftarrow \LCP (P [7..12], T [1..12]) = 6$ \\[.5ex]
line &  6: & we report $P [7..12]$ \\[.5ex]
line &  7: & $i + f - 1 = 12$ \\[.5ex]
line &  8: & we break
\end{tabular}

\vspace{5ex}

\begin{tabular}{rrrrrrrrrrrrr}
        &     1 &     2 &     3 &     4 &     5 &     6 &     7 &     8 &     9 &    10 &    11 &    12 \\
\hline
$T =$   & \tt G & \tt A & \tt T & \tt T & \tt A & \tt G & \tt A & \tt T & \tt A & \tt C & \tt A & \tt T \\
$P =$   & \tt T & \tt A & \tt C & \tt A & \tt T & \tt A & \tt G & \tt A & \tt T & \tt T & \tt A & \tt G \\
\hline
$\MF =$ &     8 &     9 &    10 &     7 &     4 &     5 &     1 &     2 &     3 &     4 &     5 &     1 \\
$\MB =$ &     3 &     5 &    10 &    11 &    12 &     9 &     6 &     7 &     8 &     4 &     5 &     6
\end{tabular}

\medskip

\caption{A trace {\bf (top)} of how Algorithm~\ref{alg:BF} processes our example {\bf (bottom)} of $P = \mathtt{TACATAGATTAG}$ and $T = \mathtt{GATTAGATACAT}$ from Figure~\ref{fig:MFMB}, with $L = 4$.}
\label{fig:trace}
\end{center}
\end{figure}

\begin{figure}[t]
\begin{center}
\resizebox{.7\textwidth}{!}{\includegraphics{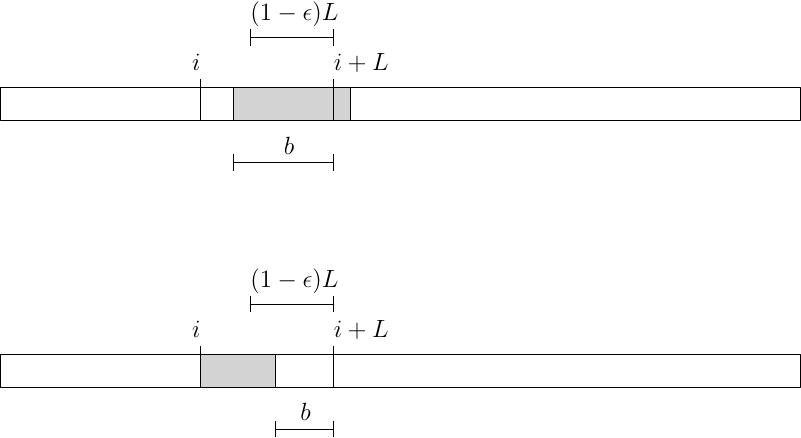}}
\caption{The two subcases of second cases ($b < L$ in line 4 of Algorithm~\ref{alg:BF}.  When $(1 - \epsilon) L \leq b < L$ {\bf (top)}, there is a MEM of length at least $L$ starting at $i_k + L - b$ {\bf (shown in grey)}.  When $b < (1 - \epsilon) L$ {\bf (bottom}), there are $L - b > \epsilon L$ characters between $i$ and $i + L - b$ {\bf (shown in grey)}.}
\label{fig:subcases}
\end{center}
\end{figure}

We can charge the $O (t (n))$ time we spend in each first case ($b \geq L$ in line 4 of Algorithm~\ref{alg:BF}) to the MEM $P [i_k..i_k + f - 1]$ that we then report, and get a bound of $O (\mu_L t (n))$ total time for all the first cases, where $\mu_L$ is the number of MEMs of length at least $L$.  To bound the time we spend on the second cases ($b < L$ in line 4 of Algorithm~\ref{alg:BF}), we observe that for each second case, we either find a MEM of length at least $(1 - \epsilon) L$ --- which may or may not be of length at least $L$, and so which we may or may not report later in a first case --- or we advance at least $\epsilon L$ characters.  These two subcases are illustrated in Figure~\ref{fig:subcases}.

Choose $\epsilon$ strictly between 0 and 1 and consider that when $(1 - \epsilon) L \leq b < L$, we can charge the $O (t (n))$ time for the second case to the MEM starting at $i_{k + 1} = i_k + L - b$, which has length at least $b \geq (1 - \epsilon) L$.  On the other hand, when $b < (1 - \epsilon) L$ we can charge a $(\frac{1}{\epsilon L})$-fraction of the $O (t (n))$ time for the second case to each of the
\[i_{k + 1} - i_k = L - b > \epsilon L\]
characters in $P [i_k..i_{k + 1} - 1]$.

Although we may charge the $O (t (n))$ time for a second case to a MEM of length at least $(1 - \epsilon) L$, and then right after charge the $O (t (n))$ time for a first case to the same MEM --- because it also has length at least $L$ --- we do this at most once to each such MEM.  In total we still charge $O (t (n))$ time to each MEM of length at least $(1 - \epsilon) L$ and $O \left( \frac{t (n)}{\epsilon L} \right)$ time to each character in $P$.  This means we use $O \left( \left( \frac{m}{\epsilon L} + \mu_{(1 - \epsilon) L} \right) t (n) \right)$ time overall, where $\mu_{(1 - \epsilon) L} \geq \mu_L$ is the number of MEMs of length at least $(1 - \epsilon) L$.  Since our algorithm does not depend on $\epsilon$, this bound holds for all $\epsilon$ strictly between 0 and 1 simultaneously.

\begin{theorem}
\label{thm:result}
Suppose we have $\MF$ and $\MB$ and we can compute in $O (t (n))$ time $\LCP \left( \rule{0ex}{2ex} P [i..m], T [\MF [i]..n] \right)$ and $\LCS \left( \rule{0ex}{2ex} P [1..i], T [1..\MB [i]] \right)$ for any $i$.  Then we can find all MEMs of $P$ with respect to $T$ with length at least a given threshold $L$ in $O \left( \left( \frac{m}{\epsilon L} + \mu_{(1 - \epsilon) L} \right) t (n) \right)$ time for all $\epsilon$ strictly between 0 and 1 simultaneously.
\end{theorem}

Combining this result with those from Section~\ref{sec:previous} gives us something like Theorem~\ref{thm:all} but with the query time depending on $\mu_{(1 - \epsilon) L}$ instead of on $\mu$.  We note that our final result does not depend on the number of MEMs much shorter than $L$ (the weakness of Theorem~\ref{thm:all}), nor on the length of MEMs much longer than $L$ (the weakness of Goga et al.'s approach).

\begin{theorem}
\label{thm:final}
Suppose $\sigma$ is polylogarithmic in $n$, $\epsilon$ is a constant strictly between 0 and 1 and $L \in \Omega (\log n)$.  Then there is an $O (r + \bar{r} + g)$-space index for $T$ with which, given $P$, we can find all the MEMs of $P$ with respect to $T$ with length at least $L$ correctly with high probability and in $O (m + \mu_{(1 - \epsilon) L} \log n)$ time. 
\end{theorem}

\section{Acknowledgments}

This work was done while the author visited Paola Bonizzoni's group at the University of Milano-Bicocca.  Many thanks to them and Lore Depuydt for helpful discussions --- especially to Luca Denti for pointing out Li's forward-backward algorithm --- and to the anonymous reviewers for pointing out some mistakes in the submitted draft and for suggesting improvements to the presentation.  This research was funded by NSERC Discovery Grant RGPIN-07185-2020 to the author and NIH grant R01HG011392 to Ben Langmead.


\newpage
\appendix

\section{Experiments}
\label{app:experiments}

Although we have not fully implemented Algorithm~\ref{alg:BF}, we have implemented a simplified version of it --- Algorithm~\ref{alg:BML} --- that uses only backward stepping instead of querying a grammar.  It may be slower in some cases, but this simplified version is deterministic and can return the BWT intervals of the MEMs as it finds them.  Our code is available on request.

\begin{algorithm}[t]
\caption{Pseudocode for a simplified and deterministic version of Algorithm~\ref{alg:BF} that uses only backward stepping.}
\label{alg:BML}
\begin{algorithmic}[1]
\State{$i \gets 0$}
\While{$i \leq m - L$}
  \State{$j \gets i + L - 1$}
  \State{$k \gets j - \search (\FMT, P, j + 1) + 1$}
  \If{$k > i$}
    \State{$i \gets k$}
  \Else
    \State{$j \gets i + \search (\FMrevT, \revP, m - i) - 1$}
	\State{{\bf report} $P [i..j]$}
    \If{$j < m - 1$}
      \State{$i \gets j - \search (\FMT, P, j + 2) + 2$}
    \Else
      \State{\bf break}
    \EndIf
  \EndIf
\EndWhile
\end{algorithmic}
\end{algorithm}

To keep the pseudocode closer to the real code, now we index strings from 0.  We use one FM-index $\FMT$ for $T$, with which we search for substrings of $P$, and another $\FMrevT$ for the reverse of $T$, with which we search for substrings of the reverse $\revP$ of $P$.  We denote a backward search for the prefix of length $\ell$ of a pattern $Q$ with an FM-index for a string $S$ as $\search (\mathrm{FMS}, Q, \ell)$, and assume it returns the number of characters matched.

We pseudo-randomly generated a binary string $T$ of ten million bits and indexed it as cyclic, then generated $P$ by flipping each of the first ten thousand bits of $T$ with probability 10\%.  (The cyclicity removes discontinuitites and makes the complexities more obvious.)  We used Li's original forward-backward algorithm to find all the MEMs of $P$ with respect to $T$ and recorded for each length how many MEMs there were of that length, how many of them occurred only once in $T$, and how many occurred only once in $T$ and in the same positions in $P$ and $T$.

The forward-backward algorithm used 188\,825 backward steps and Table~\ref{tab:experiments} shows our results.  As expected, most MEMs are short and uninformative, but most long MEMs are informative.  Algorithm~\ref{alg:BML} with $L = 40$ found the same MEMs of length at least 40 (the bottom 9 rows of Table~\ref{tab:experiments}) and used only 16\,505 backward steps.  We conjecture that at least for this experiment, the forward-backward algorithm's expected complexity is $\Theta (m \log n)$ and Algorithm~\ref{alg:BML}'s is $O (m)$, but we leave further experiments as future work.

\begin{table}[t!]
\begin{center}
\caption{For each distinct MEM length we observed in our experiment, we recorded the number of MEMs there were of that length (``count''), how many of them occurred only once in $T$ (``unique''), and and how many occurred only once in $T$ and in the same positions in $P$ and $T$ (``correct'').  As expected, most MEMs are short and uninformative, but most long MEMs are informative.}
\label{tab:experiments}
\begin{tabular}{r@{\hspace{5ex}}r@{\hspace{5ex}}r@{\hspace{5ex}}r}
length & count & unique & correct \\
\hline
20 &    5 &     0 &     0 \\
21 &    173 &   79 &    0 \\
22 &    671 &   471 &   2 \\
23 &    941 &   792 &   0 \\
24 &    852 &   790 &   5 \\
25 &    539 &   515 &   6 \\
26 &    317 &   307 &   8 \\
27 &    166 &   164 &   3 \\
28 &    90 &    90 &    6 \\
29 &    52 &    52 &    8 \\
30 &    29 &    29 &    6 \\
31 &    15 &    15 &    4 \\
32 &    7 &     7 &     1 \\
33 &    3 &     3 &     1 \\
34 &    4 &     4 &     1 \\
35 &    2 &     2 &     1 \\
36 &    2 &     2 &     2 \\
37 &    1 &     1 &     1 \\
38 &    3 &     3 &     3 \\
39 &    3 &     3 &     3 \\
42 &    1 &     1 &     1 \\
43 &    1 &     1 &     1 \\
44 &    1 &     1 &     1 \\
46 &    1 &     1 &     1 \\
50 &    2 &     2 &     2 \\
51 &    1 &     1 &     1 \\
53 &    2 &     2 &     2 \\
62 &    1 &     1 &     1 \\
74 &    1 &     1 &     1
\end{tabular}
\end{center}
\end{table}

Out of curiousity, we adjusted our code to find quickly only a longest common substring of $P$ and $T$ --- that is, a maximum-length MEM --- by keeping $L$ exactly 1 more than the length of the longest MEM found so far.  Algorithm~\ref{alg:BML} then used 6556 backward steps, and increasing then length $m$ of $P$ from ten thousand to ten million only increased the number of backward steps to 2\,462\,024 (with the length of the longest common substring increasing from 74 to 141), suggesting the complexity is $o (m)$.

\end{document}